\documentstyle[twoside,fleqn,espcrc2,epsf,rotate,epsfig]{article}

\newcommand{\chupiii}{$\chi U\phi_3$}
\newcommand{\cbcex}{\langle\overline{\chi}\chi\rangle}

\newcommand{\AmS}{{\protect\the\textfont2
  A\kern-.1667em\lower.5ex\hbox{M}\kern-.125emS}}

\title{Analysis of the Lee-Yang zeros in
a dynamical mass generation model in three dimensions} 
\author{
I.M. Barbour\address{Dept. of Physics and Astronomy,University of Glasgow G12 8QQ, U.K.,(UKQCD Collaboration)},
W. Franzki\address{Institute of Theoretical Physics E, RWTH Aachen, D-52056 Aachen,GERMANY}
and 
N.Psycharis$^{{\footnotesize{a}}}$
\thanks{Talk presented by N.Psycharis}
}       
\begin{document}

\begin{abstract}
We investigate a strongly U($1$) gauge theory with fermions
and scalars on a three dimensional lattice
and analyze whether the continuum limit might be 
a renormalizable theory with
dynamical mass generation.
Most attention is paid to the weak coupling region
where a possible new dynamical mass generation mechanism
might exist.
There we investigate the mass of the composite fermion,
the chiral condensate and 
the scaling of the Lee-Yang zeros.
 
\end{abstract}

\maketitle
\section{Introduction}
The fermion-gauge-scalar model ($\chi U\phi$ model),
has been suggested as a model for dynamical mass
generation in four dimensions in \cite{new2,new3}.
Investigating such a lattice model in three dimensions
\cite{new1} we find three
regions in the $\beta-\kappa$ plane with possibly different critical
behaviour
in the chiral limit of the model. These three regions are presented in
Fig.($1$).
The region X of the phase diagram
is conceivably  analytically connected with either the Nambu or
Higgs phase but it may well be
a separate phase. 
In this region the measured fermion mass is large but the
chiral condensate is very small (zero within our numerical accuracy).
We also investigated  the scaling of the 
Lee-Yang zeros in
this interesting  region of the phase diagram
in order to clarify the situation.
\begin{figure}[htb]
\epsfig{file=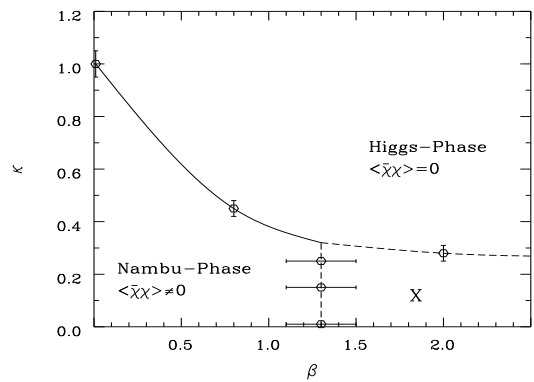,width=6cm,height=10cm,angle=90}
\vspace{-1.6cm}
\caption{Phase diagram of the \chupiii\ model for $m_0=0$. 
       All phase transitions seem to be 2$^{\rm nd}$
      order.} 
\label{fig:pd}
\vspace{-0.7cm}
\end{figure}

\section{The model}
The \chupiii\ model is defined by the lattice action:
\begin{equation}
  \label{chupact}
  S_{\chi U \phi} = S_\chi + S_U + S_\phi,
\end{equation}
with
\begin{eqnarray*}
  S_\chi &=& \frac{1}{2} \sum_x \overline{\chi}_x
  \sum_{\mu=1}^3 \eta_{\mu x} (U_{x,\mu} \chi_{x+\mu} - U^\dagger_{x-\mu,\mu}
  \chi_{x-\mu})\; \\
  &+&am_0 \sum_x \overline{\chi}_x \chi_x \;,\\
  S_U &=& \beta \sum_{x,\mu<\nu} (1-{\rm Re}\,{U_{x,\mu\nu}}) \;,\\
  S_\phi &=& - \kappa \sum_x \sum_{\mu=1}^3
  (\phi^\dagger_x U_{x,\mu} \phi_{x+\mu} + {\rm h.\,c.}) \;.
\end{eqnarray*}
$\chi_x$ are the Kogut-Susskind fermion fields. 
We
are interested in the chiral limit $m_0=0$.

\subsection{Observables}

In order to investigate the chiral properties 
of the model we concentrate first on  two observables:
the chiral condensate and the fermion mass\cite{new1}.

The third observable we concentrate on 
are the Lee-Yang zeros.
The canonical partition function of the system
can be written as\cite{new4,new1}:
\begin{eqnarray}
  Z(\beta,\kappa,am_{0})  
  \label{eq:scalim}
  &=& \sum_{n=0}^{V \over 2} A_n[\beta,\kappa] (am_{0})^{2n}.
\end{eqnarray}
The Lee-Yang zeros are the zeros of this polynomial representation of the
partition function.
The errors in the Lee-Yang zeros are estimated by a Jacknife method.

The critical properties of the system are determined by the zeros lying
closest to the real axis. 
$y_1$ is the zero with the smallest imaginary part
and is called edge singularity.
With increasing finite volume it
converges to the critical point.  For a continuous phase transition the
position of the zeros closest to the real axis in the complex plane is ruled
by the scaling law
\begin{equation}
  y_{i}(\beta,\kappa,L) - y_{R}(\beta,\kappa,\infty) = A_i L^{-1/s}
\end{equation}
where $A_i$'s are complex numbers
 and the exponent $s=s(\beta,\kappa)$\cite{new1} describes
the finite size scaling of the correlation length.
It follows that the real and the imaginary part of the zeros 
should scale
independently with the same exponent.

This scaling law
can also be extended to the case of  a first order phase
transition.  
In this case, for a
three-dimensional model we expect $s=\frac{1}{3}$.

Far away from the
critical point we expect linear scaling\cite{new1} in the
log-log plot with $s=1/3$
in the 
broken phase
and $s=1$ in the symmetric phase.
At the critical point itself we expect linear scaling.

\section{Universality at Strong Coupling}
An investigation of the scaling of the 
Lee-Yang zeros, the chiral condensate and the fermion
mass
at $\beta=0$ and $\beta=0.80$,
presented in\cite{new1},
shows that the chiral phase transition of
the \chupiii\ model is in the same universality class
at both these $\beta$ values.
At $\beta=0$ the
scalar and gauge fields can be integrated out exactly and we end up with a
lattice version of the GN$_3$ model. This model is known to have a chiral
phase
transition which is nonperturbatively renormalizable.

It is also very likely that the same universality class
holds also for $0\leq \beta\leq0.80$.
 
\section{Weak coupling region}
We first look at the neutral fermion mass
and the chiral condensate in the weak
coupling region.

The mass of the neutral fermion is
large for $\kappa < 0.27$
and it becomes 
small
for larger $\kappa$ at $\beta=2.0$\cite{new1}. 
For $\kappa>0.27$ its value probably vanishes  in the infinite
lattice size limit.

The condensate show a crossover behaviour at $\kappa\approx0.27$ with
$$\cbcex(am_0)_{\kappa<0.27} < \cbcex(am_0)_{\kappa>0.27}$$
also at $\beta=2.0$\cite{new1}. 
However, at large
$\kappa$ we believe that we are in the Higgs phase where the condensate is
zero in the chiral limit.
Therefore,in this limit,
the condensate seems to be  
zero for all $\kappa$
at $\beta=2.0$. 

At $\kappa=0.25$ and at various $\beta$,
the condensate is large at small $\beta$
but becomes very small(zero) in the chiral limit for
$\beta > 1.5$.
The fermion mass decreases for increasing $\beta$ but then
stabilizes
with $am_F>1$. So it is clearly non zero at all $\beta$.

It is surprising but also interesting 
that the fermion mass is different from
zero with unbroken
chiral symmetry in the region X 
of the phase diagram.

\subsection{The Lee-Yang zeros at Weak Coupling}
In order to clarify the situation in
the X region of the phase diagram we
investigated the Lee-Yang zeros.

The edge singularity $y_1$
has a nonzero real part
in this region compare to  the other regions of the phase diagram
where the zeros have  zero real part.
The zeros appear in conjugate pairs. We define the edge singularity
in this region to be the zero with smallest positive imaginary
part and count each pair only once.

\begin{figure}[htb]
\epsfig{file=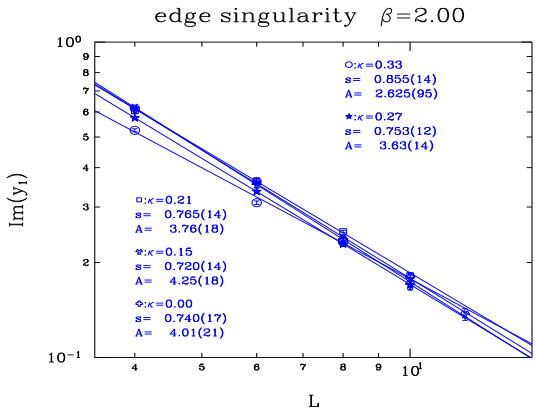,width=6cm,height=10cm,angle=90}
\vspace{-2.0cm}
\caption{Imaginary part of $y_1$ as function of the lattice size
         for different $\kappa$ at $\beta=2.00$.}
\label{fig:lyk000}
\vspace{-1.0cm}
\end{figure}

Fig.~\ref{fig:lyk000} shows the behaviour of the imaginary part
of the edge singularity as a function of lattice size at
$\beta=2.00$ and  various  $\kappa$. 
We see that the exponent $s$ seems to be  universal 
for $\beta=2.00$ and small $\kappa$ values,
with a value of $s\approx0.7$.

\begin{figure}[htb]
\epsfig{file=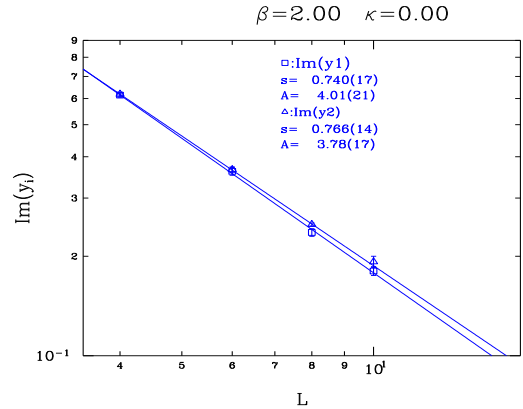,width=6cm,height=10cm,angle=90}
\vspace{-2.0cm}
\caption{Imaginary part of the first two Lee Yang zeros
          for $\beta=2.00$, $\kappa=0.00$ as
          function of lattice size.}
\label{fig:lyk001}
\vspace{-1.0cm}
\end{figure}

\begin{figure}[htb]
\epsfig{file=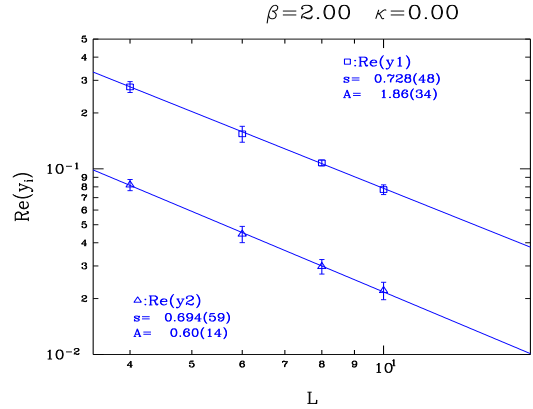,width=6cm,height=10cm,angle=90}
\vspace{-1.6cm}
\caption{Real part of the first two Lee Yang zeros
          for $\beta=2.00$, $\kappa=0.00$ as
          function of lattice size.}
\label{fig:lyk002}
\vspace{-0.6cm}
\end{figure}

Figs~\ref{fig:lyk001} and ~\ref{fig:lyk002}
show the finite size scaling of the 
imaginary and the real part of the lowest zeros
for $\beta=2.00$ and  $\kappa=0.00$, respectively.
This is  a point in the 
the region X.
The first two zeros have, within the numerical precision,
identical imaginary part but their real parts differ by a factor of about
3.5.
Their imaginary  
parts scale linearly in the log-log plot with the same exponent
s within the numerical precision.
Although with somewhat larger errors,
the same seems to happen also for their real parts.
This pattern of zeros can be 
found
everywhere in the region X.

%
\section{Conclusions}

The results in the weak coupling region,
and specially in the interesting X region,
show that the X region could belong to the
Nambu phase if the condensate is
really nonzero but unmeasurably small.
It could also belong to the
Higgs  phase if
the charged fermion becomes massless in the chiral
limit and is no longer a bound state.
Finally there is also the explanation that it
could be a separate phase
with a new dynamical mass generation mechanism.
The Lee-Yang zeros now distinguish the region X
from the Nambu and Higgs phase by a
nonvanishing real part of the lowest zeros
and a scaling, which can
neither
can be described by an exponent $s=1/3$ nor $s=1$.
%

\section{Acknowledgements}
        We would like to thank E. Focht,J. Jers{\'a}k and
        M.A. Senechal for very useful suggestions.
\

\end{document}